\documentclass{llncs}

\usepackage[table]{xcolor}
\usepackage{array}
\usepackage{tabu}   
\usepackage{xspace} 

\usepackage{comment}

\usepackage{graphicx}

\usepackage{hyperref}

\usepackage{color}
\usepackage{xcolor}

\usepackage{todonotes}

\usepackage{tikz}
\usepackage{tikz-qtree}

\usepackage{etex}
\usepackage{booktabs}
\usepackage{pgfplotstable}
\usepackage{adjustbox}
\usepackage{todonotes}

\begin{document}

\title{A User Evaluation of \\Automated Process Discovery Algorithms}
\titlerunning{Hamiltonian Mechanics}  
%
\author{Fabrizio Maria Maggi\inst{1} \and Andrea Marrella\inst{2}
\and Fredrik Milani\inst{1} \and \\ Allar Soo\inst{1} \and Silva Kasela\inst{1}}
\authorrunning{Maggi et al.} 
%
%
\institute{University of Tartu, Tartu, Estonia,\\ \email{\{f.m.maggi,milani,soo,silva.kasela\}@ut.ee} \and
Sapienza Universit\'{a} di Roma, Rome, Italy\\
\email{marrella@diag.uniroma1.it}
}

\maketitle              

\begin{abstract}
Process mining methods allow analysts to use logs of historical executions of business processes in order to gain knowledge about the actual behavior of these processes. One of the most widely studied process mining operations is automated process discovery. An event log is taken as input by an automated process discovery method and produces a business process model as output that captures the control-flow relations between tasks that are described by the event log. In this setting, this paper provides a systematic comparative evaluation of existing implementations of automated process discovery methods with domain experts by using a real-life event log extracted from an international software engineering company and four quality metrics. The evaluation results highlight gaps and unexplored trade-offs in the field and allow researchers to improve the lacks in the automated process discovery methods in terms of usability of process discovery techniques in industry.
\end{abstract}

\section{Introduction}
Today's competitive business environment combined with digital technologies pose a choice to companies, namely to improve or fade away. To improve, they need to increase the efficiencies of their business processes. For decades, analysts have relied on manually modeling the processes and analyzing them so to identify improvement opportunities.
Nowadays, by combining business process thinking and data analytics into \textit{process mining} techniques, companies are in a position to take process analysis and improvement to new levels.

According to \cite{manifesto} process mining can be divided into three main branches, process discovery \cite{DBLP:journals/tkde/AalstWM04,van2009min,mining2011discovery}, conformance checking \cite{conf_check_1,conf_check_2,conf_check_3,conf_check_4} and process enhancement \cite{predicition,predicition2,fahland2012repairing}.
Process discovery has been and remains the most common and widely studied use case \cite{mining2011discovery}. With an event log (capturing unique case ids, activities, and timestamps) as input, every process discovery method produces a business process model. Nevertheless, there are some challenges in this field. Indeed, the models generated must manage the trade-off of four metrics \cite{buijs2012role}: fitness, generalization, precision, and simplicity. Over the past decade, impressive advancements have been made in this field \cite{mining2011discovery}. Despite this, automated process discovery methods suffer from two recurrent deficiencies when applied to real-life logs \cite{de2012multi}: (i) they produce large and spaghetti-like models; and (ii) they produce models that do not manage to find the right trade-off of the four metrics mentioned above. If such models are difficult to understand or perceived as imprecise, they fail to become the valuable tool they are designed to be.

The evaluation of discovery algorithms is done by using logs (most commonly real-life industry logs) where the generated models are assessed using different metrics. However, the models are rarely evaluated by the process participants or domain experts. In light of this, we seek to investigate how domain experts view and perceive process mining algorithms for process discovery. The research question of this paper is therefore ``\textit{which discovery algorithm is perceived as the best one by domain experts?}''. In so doing, we conduct a systematic literature review aiming at identifying the main discovery algorithms available today. The last overview of this field was conducted by De Weerdt et.al.~\cite{de2012multi} in 2012. We take their results as input and focus on research published after 2012. We then apply the selected algorithms on an issue tracking system from the software development department of an ERP company. The models produced are then evaluated using surveys and interviews with the domain experts of that company. As such, the contribution of this paper is (\textit{i}) an updated overview of discovery algorithms, and (\textit{ii}) a comparative evaluation of discovery algorithms from the perspective of domain experts.

The rest of the paper is structured as follows. Section \ref{sec:review} presents the systematic literature review methodology and classifies the approaches identified. Section \ref{sec:evaluation}  discusses the set up and method of evaluation while Section \ref{sec:results} discusses the evaluation and its results. Finally, Section \ref{sec:conclusion} concludes the paper and outlines future work directions. 
\section{Systematic Literature Review}
\label{sec:review}

In order to identify and analyze relevant studies (and underlying methods) related to automated (business) process discovery from event logs, we conducted a \emph{Systematic Literature Review} (SLR) through a scientific, rigorous and replicable approach as specified by Kitchenham~\cite{Kitchenham}. For the sake of space, in this paper we will briefly present the major steps of our SLR and the most interesting results derived from its enactment. Interested readers can refer to \cite{SLR-TKDE} for a detailed discussion of the SLR.

\smallskip
\noindent
\textbf{(1) Formulation of the research questions.} Since the goal of the SLR was to select methods that produce process models from event logs, we scoped the search by formulating five research questions aimed at: \emph{(i)} identifying existing studies proposing methods to perform automated process discovery; \emph{(ii)} categorizing the output of a method on the basis of the type of process model discovered (i.e., imperative, declarative or hybrid), and the specific language employed (e.g., Petri nets, BPMN, Declare); \emph{(iii)} delving into the specific language constructs supported by a method (e.g., exclusive choice, parallelism, loops); \emph{(iv)} exploring what tool support the different methods have; and \emph{(v)} investigating how the methods have been evaluated and in which application domains.

\smallskip
\noindent
\textbf{(2) Search strings development and data sources selection.} Next, we developed four search strings by building combinations of keywords derived from our knowledge of the subject matter. The keywords employed to perform the search were: \emph{(i)} ``process discovery''; \emph{(ii)} ``workflow discovery''; \emph{(iii)} ``process learning''; and \emph{(iv)} ``workflow learning''. We intentionally excluded the term ``automated'' in the search strings, because it is often not explicitly used and led to retrieving many more studies than those that actually focus on automated process discovery.
We applied each of the search strings to seven popular academic databases: Scopus, Web of Science, IEEE
Xplore, ACM Digital Library, SpringerLink, ScienceDirect and Google Scholar, and retrieved studies based on the
occurrence of one of the search strings in the title, the keywords or the abstract of a paper. If a query on a specific data source returned more than one thousand results, we refined it by combining the search string with the term ``business'' or ``process mining'' to obtain more focused results, e.g., ``process discovery AND process mining'' The search was completed in December 2017.

\smallskip
\noindent
\textbf{(3) Definition of inclusion criteria.} Then, we defined inclusion criteria to draw the borders of our search scope and ensure an unbiased selection of relevant studies. To be selected, a study must satisfy all the inclusion criteria, which allowed us to retain studies that: \emph{(i)} propose a method for automated process discovery from event logs; \emph{(ii)} propose a method that has been implemented and evaluated; \emph{(iii)} are peer-reviewed, written in English and published in 2011 or later (earlier studies have been evaluated by De Weerdt et al. in \cite{BrouckeWVB13}).

\smallskip
\noindent
\textbf{(4) Study selection.} Finally, we used the search strings to conduct a search on the selected data sources. By only applying the last inclusion criterion, the search queries initially yielded a total of 2820 potentially relevant studies. Then, we analyzed title, abstract, introduction, conclusion and evaluation of these studies to exclude those studies that were clearly not compliant with the other inclusion criteria. As result of the iterations, we found 86 studies matching all the inclusion criteria. However, many of these studies refer to the same automated process discovery method, i.e., some studies are either extensions, optimization, preliminaries or generalization of another study. For such reason, we decided to group the studies by either the last version or the most general one. At the end of this process, 35 main groups of discovery methods were identified.

\smallskip
\noindent
\textbf{(5) Study classification.}
Driven by the research questions, we also classified the methods underlying these studies on the basis of the following dimensions: \emph{(i)} model type (procedural, declarative, hybrid) and language (e.g., Petri nets, BPMN, Declare) supported; \emph{(ii)} semantics captured in procedural models, e.g., parallelism (AND), exclusive choice (XOR), inclusive choice (OR) and loop; \emph{(iii)} type of implementation (standalone or plug-in, and tool accessibility); \emph{(iv)} type of evaluation data (real-life, synthetic or artificial log) and \emph{(v)} domain of application (e.g., insurance, banking, healthcare, etc.). 
Collectively, this information is summarized in Table 2 of work \cite{SLR-TKDE}, where each entry refers to the main study of the 35 groups found.


\section{Evaluation}
\label{sec:evaluation}
In this section, we describe the set up and method of our evaluation. In section \ref{sec:used_log}, we give a general description of the log used and specify the list of miners used. In section \ref{sec:mined_models}, we describe the preprocessing applied to the original log to create a refined the dataset to be used for model discovery. In section \ref{sec:evaluation_process}, we specify the setup for the user evaluation. In section \ref{sec:statistical_analysis}, we present the methods for conducting our statistical analysis. 

\subsection{Experimental Setting}
\label{sec:used_log}
In our evaluation, we use a log from a software company pertaining to a software development process for developing an ERP system. The log contains data spanning over one year. It has $52\,629$ events and $5\,551$ cases. The log is extracted from their own software used for functional enhancement and bugs. The input log does not contain explicit information about activities performed. Therefore, in section \ref{sec:mined_models}, we describe how this information was extracted from the log so to make it suitable for process discovery. After applying this procedure, 29 unique activities were identified.

In the evaluation, we used a selection of the methods surveyed in \cite{SLR-TKDE}.
Assessing all the methods that resulted from the literature review would not be possible due to the heterogeneous nature of the inputs required and the outputs produced. Hence, we decided to focus on the largest subset of comparable methods. The methods considered were the ones satisfying the following criteria:
\begin{itemize}
\item an implementation of the method is publicly accessible;
\item the output of the method is a BPMN model or a model seamlessly convertible into BPMN (i.e., process trees and Petri nets).
\end{itemize}
The second criterion is a requirement dictated by the fact that evaluation was performed with business users from industry, which are often non-expert of the technical base formalisms used in the BPM field.
Using these criteria, the following miners were identified:
\begin{itemize}
\item BPMNMiner \cite{conforti2016bpmn}, which is a method for the automated discovery of BPMN models containing sub-processes, activity markers such as multi-instance and loops, and interrupting and non-interrupting boundary events (to model exception handling).
The method is robust to noise in event logs.
\item Causal Net Mining \cite{greco2015process}, which encodes causal relations gathered from an event log and if available, background knowledge in terms of precedence constraints over the topology of the resulting process models. The discovery algorithm is formulated in terms of reasoning problems over precedence constraints.
\item alpha\$-algorithm \cite{guo2015mining}, which can discover invisible tasks involved in non-free-choice constructs. The algorithm is an extension of the well-known $\alpha$ algorithm, one of the very first algorithms for automated process discovery, originally presented in~\cite{DBLP:journals/tkde/AalstWM04}.
\item Heuristics Miner \cite{weijters2011flexible}, which is a method that can discover process models containing non-trivial constructs, but with a low degree of block-structuredness. At the same time, the method can cope well with noise in event logs.
\item Hybrid ILP Miner \cite{van2015ilp}, which is based on hybrid variable-based regions. Through hybrid variable-based regions, it is possible to vary the number of variables used within the ILP (Integer Linear Programming) problems being solved. Using a different number of variables has an impact on the average computation time for solving the ILP problem.
\item Structured Miner \cite{augusto2017automated}, which is an improvement of the Heuristics Miner algorithm to separate the objective of producing accurate models and that of ensuring their structuredness and soundness. Instead of directly discovering a structured process model, the approach first discovers accurate, possibly unstructured (and unsound) process models, and then transforms the resulting process model into a structured (and sound) one.
\item Inductive Miner \cite{leemans2014discovering}, which is based on the extraction of process trees from an event log. It efficiently drops infrequent behavior, still ensuring that the discovered model is behaviorally correct (sound) and highly fitting.
\item Evolutionary Tree Miner (ETMd) \cite{buijs2014quality}, which is
based on a genetic algorithm that allows the user to drive the discovery process based on preferences with respect to the four quality dimensions of the discovered model: fitness, precision, generalization and complexity.
\end{itemize}


\subsection{From the event log to the process models}
\label{sec:mined_models}
As mentioned in section \ref{sec:used_log}, the log was preprocessed for the evaluation. 
The activities were not explicitly recorded in the log but the information about the actors/process participants was.
However, the company from where the log comes, has an organizational structure where one department performs quality assurance and another documentation. By using the role of the actors, such as \textit{EE Senior Coder}, the activity could be deduced. This step was conducted together with the company to ensure that the activities were correctly captured. 

\begin{table}[t!]
\centering
\scriptsize
\caption{Complexity of the discovered models before filtering}
\label{tab:unfilteredComplexity}
\begin{tabular}{c|c|c|c}
\textbf{Miner}            & \textbf{Size} & \textbf{CNC} & \textbf{Density} \\ \hline
\textbf{alpha\$}          & 145           & 1.490        & 0.010            \\ \hline
\textbf{BPMN Miner}       & 25            & 1.760        & 0.073            \\ \hline
\textbf{CNM}              &   122            & 2.155             & 0.017                 \\ \hline
\textbf{ETMd}             & 124           & 1.411        & 0.011            \\ \hline
\textbf{HILP}             & 65            & 1.600        & 0.025            \\ \hline
\textbf{HM}              & 97            & 1.990        & 0.021            \\ \hline
\textbf{IM}             & 42            & 1.571        & 0.038            \\ \hline
\textbf{Structured Miner} & 54            & 1.982        & 0.037           
\end{tabular}
\end{table}

\begin{table}[t!]
\centering
\scriptsize
\caption{Complexity of the discovered models after filtering}
\label{tab:filteredComplexity}
\begin{tabular}{c|c|c|c}
\textbf{Miner}            & \textbf{Size}              & \textbf{CNC}                  & \textbf{Density}              \\ \hline
\textbf{alpha\$}          & 71                         & 1.408                         & 0.020                         \\ \hline
\textbf{BPMN Miner}       & 15                         & 1.600                         & 0.114                         \\ \hline
\textbf{CNM}              &    52                        &   1.411                            &  0.020                              \\ \hline
\textbf{ETMd}             & 84                         & \cellcolor{gray!25}\textbf{1.274} & \cellcolor{gray!25} \textbf{0.015 }\\ \hline
\textbf{HILP}             & 34                         & 1.471                         & 0.045                         \\ \hline
\textbf{HM}              & 52                         & 1.865                         & 0.037                         \\ \hline
\textbf{IM}             & \cellcolor{gray!25}\textbf{28} & 1.429                         & 0.053                         \\ \hline
\textbf{Structured Miner} & \cellcolor{gray!25}\textbf{33} & 1.454                         & 0.045                        
\end{tabular}
\end{table}

When applying the identified methods to the original log, all the BPMN models discovered were highly complex and spaghetti-like. In \tablename~\ref{tab:unfilteredComplexity}, we show the metrics measuring the complexity of the models discovered from the original log. In particular, size is the number of nodes, CNC is the Coefficient of Network Connectivity (CNC), i.e., the ratio between the number of arcs and the number of nodes and density is the ratio between the actual number of arcs and the maximum possible number of arcs in any model with the same number of nodes. 

\begin{figure}[t!]
	\centering
	\includegraphics[width=\textwidth]{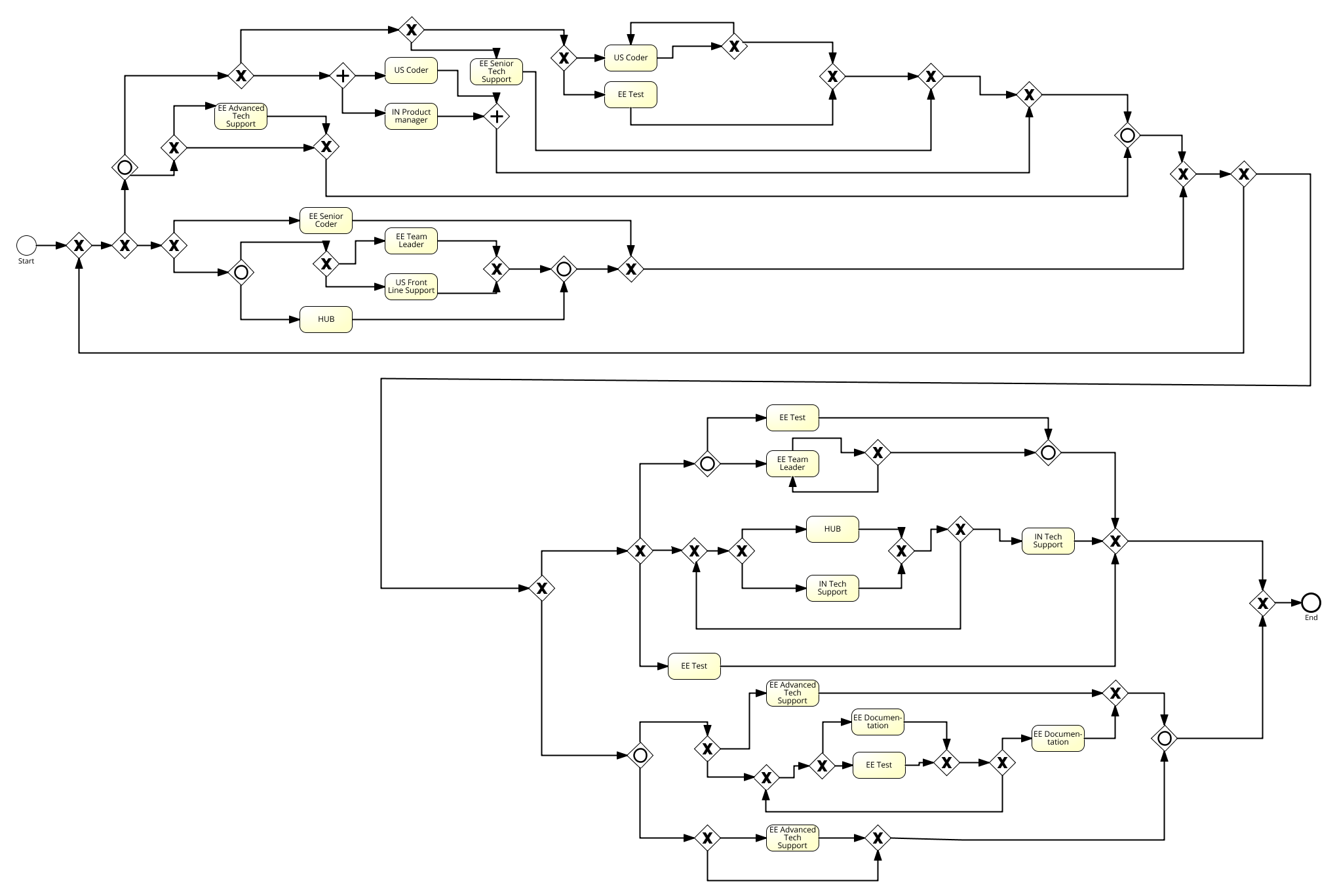}
	\caption{BPMN model mined by Evolutionary Tree Miner (model A)}
	\label{fig:grid_plotEvol}
\end{figure}

\begin{figure}[t!]
	\centering
	\includegraphics[width=\textwidth]{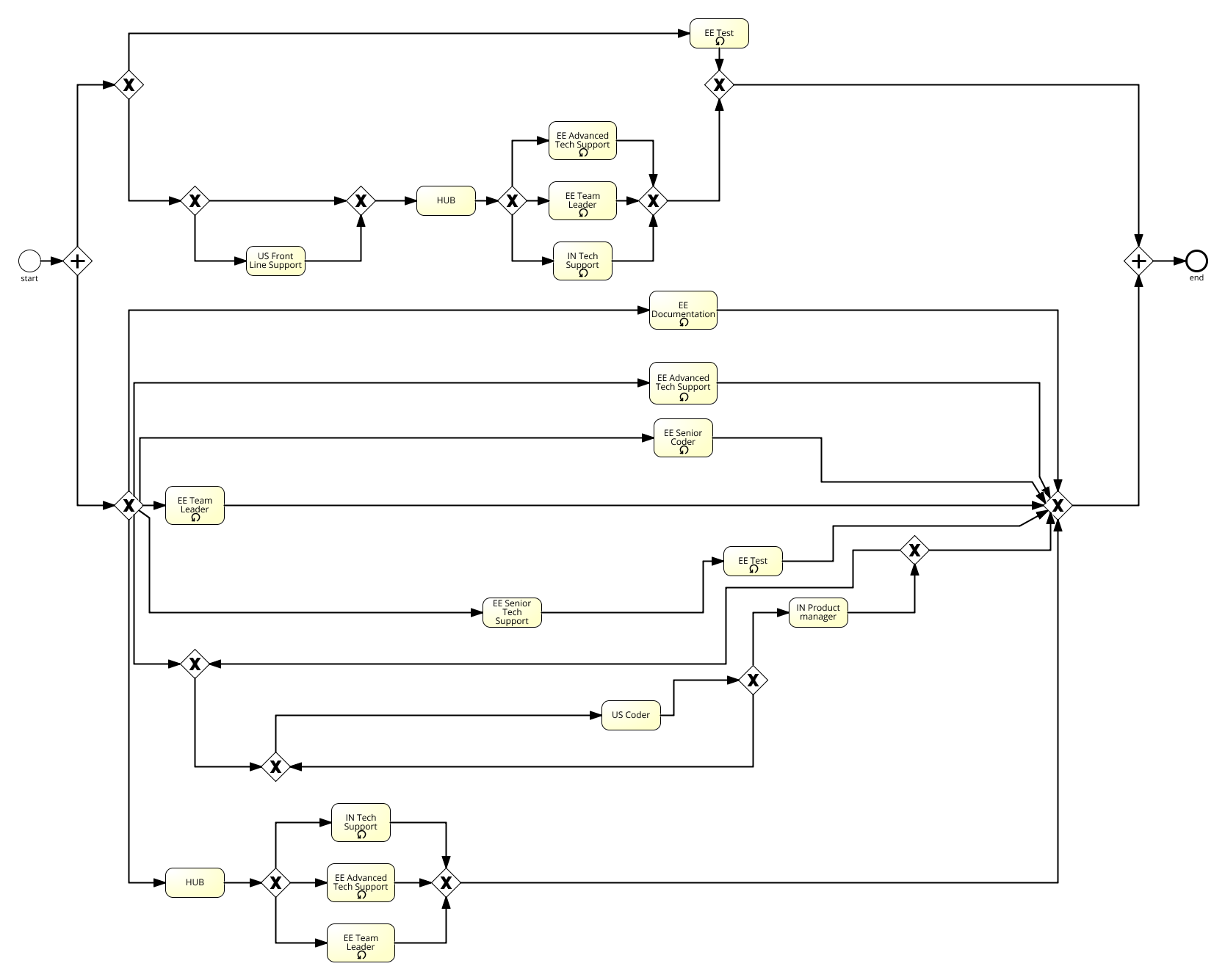}
	\caption{BPMN model mined by Structured Miner (model B)}
	\label{fig:grid_plotInd}
\end{figure}

\begin{figure}[t!]
	\centering
	\includegraphics[width=\textwidth]{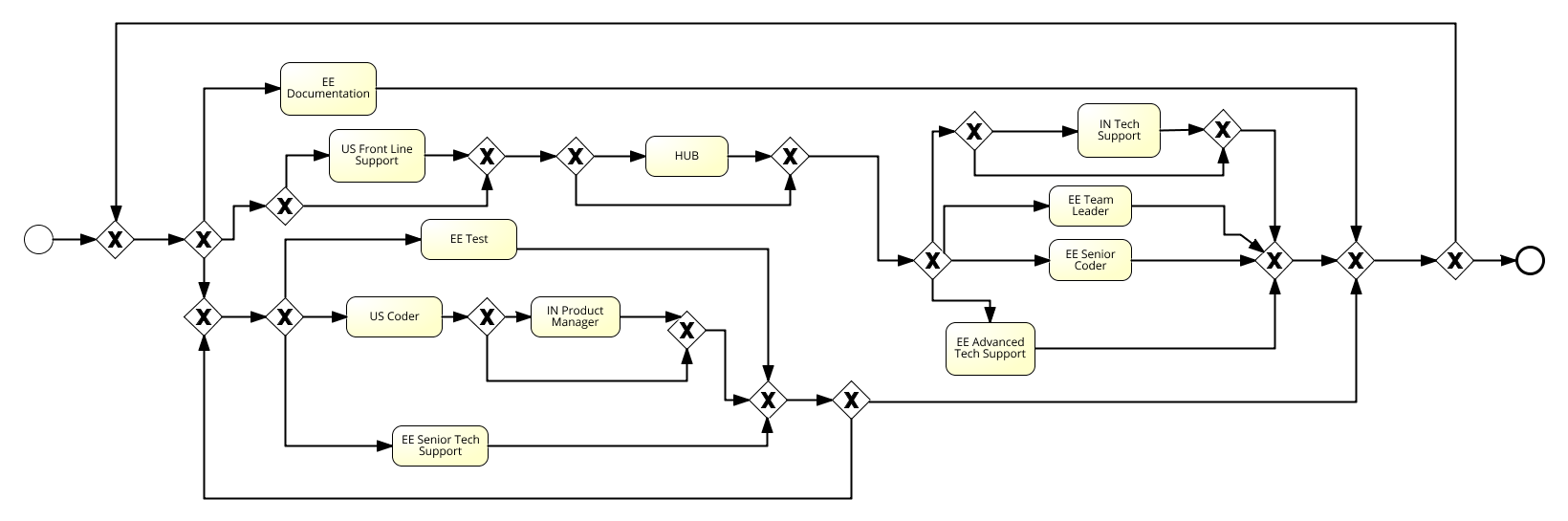}
	\caption{BPMN model mined by Inductive Miner (model C)}
	\label{fig:grid_plotStruct}
\end{figure}

To improve the understandability of the models, we decided to filter the original log to isolate frequent behaviors. In particular, we created nine separate logs ranging from a log containing all behavior, to a log containing the behavior shared by at least 2 cases up to a log containing the behavior shared by at least 9 cases. 
These logs were then used to produce a BPMN model. 
In \tablename~\ref{tab:filteredComplexity}, we show the metrics measuring the complexity of the models discovered from the log containing the behavior shared by at least 9 cases.
For the user evaluation, we selected the BPMN model mined by Evolutionary Tree Miner represented in \figurename~\ref{fig:grid_plotEvol} since this model has the lowest CNC and density\footnote{Model A produced many redundant elements which we manually removed to make the model more readable for the domain experts}. The second model chosen was the one obtained by the Structured Miner shown in \figurename~\ref{fig:grid_plotStruct} due to its smallest size. Finally, we also included the model generated by the Inductive Miner as shown in \figurename~\ref{fig:grid_plotInd}. For anonymization purposes during the evaluation, we referred to the Evolutionary Tree Miner as model A, the Structured Miner as model B, and the Inductive Miner as model C. We discarded the model discovered with BPMN Miner since, even if very simple, this model was very imprecise (a ``flower'' model allowing any behavior).

\subsection{Evaluation set-up}
\label{sec:evaluation_process}
The evaluation was conducted in two steps both of which involved the domain experts of the company. Having produced the models, we set up several meetings where employees from development teams, product managers, testers, documenters, and all team leads were invited to. In total, 18 domain experts participated in the survey which constitute 72\% of process participants at the company (excluding administrative staff). In the first step, the participants were given a printed copy of the papers and asked to fill out a questionnaire (via Google Forms). The questions asked concerned their familiarity with process models and work with such models over the past 12 months. Next, they were asked to compare the models from different process model quality metrics. The questions asked were as follows.


	\begin{enumerate}
		\item \label{DE1} Rate how easy it is for you to understand the process models (1 means very difficult, 7 means very easy).
		\item \label{DE2} Take one path and follow it from the beginning to the end. Rate how easy it is for you to follow your chosen path (1 means very difficult, 7 means very easy).
		\item \label{DE3} Rate how easy it is for you to distinguish the paths in models (1 means very difficult, 7 means very easy).
		\item \label{DE4} Can you recognize any processes you work with in the models? (1 means not at all, 7 means yes, clearly, everything is there).
		\item \label{DE5} In your estimation, rate how well the models describe your processes (1 means that the model is too specific so to exclude some paths that are possible in reality, 7 means that the model is too general so to allow process paths that are not possible in reality).
		\item \label{DE6} If you were to improve your business processes, which model would you find most useful for this purpose? (1 means useless, 7 means very useful).
	\end{enumerate}

\noindent
The above questions correspond to the following process model quality metrics:
\begin{itemize}
	\item \textbf{understandability} - Questions \ref{DE1}, \ref{DE2} and \ref{DE3};
    \item \textbf{correctness} - Question \ref{DE4};
	\item \textbf{precision} - Question \ref{DE5};
	\item \textbf{usefulness} - Question \ref{DE6}.
\end{itemize}



For each question we had three variants as A, B or C representing the specific miner used to create the model. We restricted ourselves to three variants also to avoid getting meaningless results due relativity low amount of participants. If we would have more than three models, the answers could be distributed in a way where it wouldn't be possible to make any statistically strong conclusions.

In the second stage, which was performed after first stage, we carried out a workshop. It was in open form allowing participants to discuss and express their perceptions and offer qualitative feedback about the models compared. The discussions did not follow a strict structure but we used the following targeted questions to moderate the workshop.

\begin{itemize}
	\item Which of the models were the best? Why?
	\item How did the models look like in general?
	\item What could be developed?
	\item Did the models filled your expatiations?
	\item Would you consider using these algorithms in your company? And process discovery?
	\item What lacks are present in the models?
\end{itemize} 

\subsection{Description of statistical analysis methods used}
\label{sec:statistical_analysis}
With statistical analysis we wanted to discover if there is any differences of the ratings between the models. Before the analyses, the data was formatted to be suitable for data analysis with the free software R. The answers of the questionnaire was extracted from Google Forms as .CSV file. At first the name of the questions in the header of the .CSV file were renamed to correspond to the respective metrics. If there was more than one question for a metric, the ratings of the subquestions were divided by the number of subquestions and then summed to reflect the general rating of the respective process model quality metrics.\\ \\
For the evaluation of domain experts perception of the models, we formulated the following hypothesis pair:
\begin{itemize}
	\item \textbf{The null hypothesis}: There is no difference in the mean rating of the models.
	 \item \textbf{The alternative hypothesis}: There is at least one model that is different from the others.
\end{itemize}

The hypotheses were tested using the two-way ANOVA. If we assume the independence of the observations, then the ANOVA model additionally assumes that the residuals are normally distributed for each combination of the groups and the residuals have the same variance (homogeneity of variances) for each combination of the groups. Normality assumption was assessed with Shapiro-Wilk test and QQ-plots, homogeneity of variances was assessed with Levene's test. A significant ANOVA test was followed by Tukey HSD test to perform multiple pairwise-comparison between the means to determine the statistically significant pairs of groups. All the analysis was done in R by using RStudio and the figures were made with ggplot2. Violin plots were used due their expressiveness of median value, interquartile range, and kernel density estimations.

\section{Evaluation results}
\label{sec:results}
The research question of this paper is about how discovery algorithm are perceived by domain experts. In answering this question, 
we investigate the perceived understandability, correctness, precision, and usefulness of the models by the domain experts. Size and complexity together with perceived understandability correspond to ``simplicity''. Correctness aims at assessing perceived ``fitness'' whereas precision captures perceived ``generalization'' and ``precision''. In this section, we describe the results of the evaluation from two perspectives. The first are results obtained from surveys filled in by the domain experts. Secondly, we summarize the main results gathered from discussions with the domain experts. The algorithm generating the models are referred to as model A (Evolutionary Tree Miner), model B (Structured Miner), and Model C (Inductive Miner).



\subsection{Domain Expert Survey Results}
\begin{figure}[htb!]
	\centering
	\includegraphics[width=\textwidth]{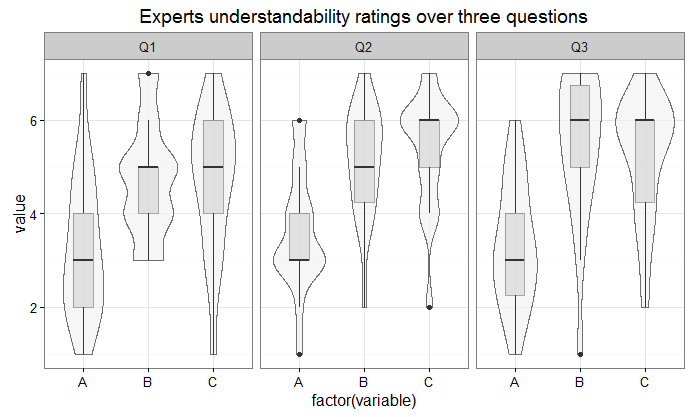}
	\caption{Results for Understandability}
	\label{fig:EQ1_plot}
\end{figure}

The domain experts were given a link to a survey where did saw the three models accompanied with 6 questions. These questions aimed at measuring the perceived understandability, precision, correctness, and usefulness. The first three questions aimed at assessing perceived understandability. These questions asked about how easy it was understand the model, to follow one path from the beginning to the end, and to distinguish between the different paths. The respondents gave a rating from 1 (very difficult) to 7 (very easy) for each of the selected models (model A, B, and C). In regards to understanding the models, A had a mean of 3.28 whereas B and C both had 4.72. As to the ease by which a path could be followed, A had the lowest mean value of 3.61, B the second lowest with 5.11 and C the highest with 5.39. The experts also found model A to be most difficult when distinguish between different paths. The mean values for this question are 3.33 for A, 5.39 for B, and 5.22 for C. The results are also depicted in Figure \ref{fig:EQ1_plot} (the bold line denotes the median). As can be seen, model A is perceived as the least understandable for all three questions whereas model B and C are comparable. The difference between model A and the others is most significant in regards to the ease by which different paths can be distinguished in the model. It is interesting to note that the respondents perceived it slightly easier to follow a path from beginning to end in model C as compared to model B.
The results for understandability seem to show a distinct preference for model B and C over model A, in particular for distinguishing the different paths in the models.

The fourth question of the survey aimed at capturing the perceived correctness of the models. This was achieved by asking the respondents if they could recognize processes they work with in the models. For each model, they gave a rating from 1 (not at all) to 7 (all of them).
\begin{figure}[htb!]
	\centering
	\includegraphics[width=\textwidth]{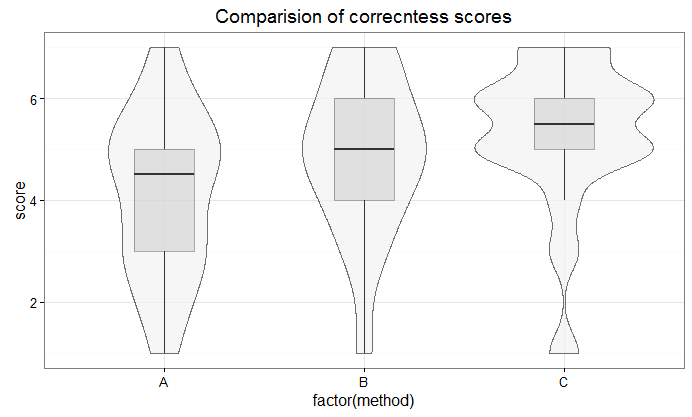}
	\caption{Results for Correctness}
	\label{fig:EQ2_plot}
\end{figure}
The results \ref{fig:EQ2_plot} show that models are quite similar in regards to perceived correctness. However, there is a slight preference for model C over B and A. The interquartile range and density for model C puts it as the model perceived to be most correct by the domain experts. This is also reflected in the mean values where model A has 4.11, B with 4.89, and finally C with 5.28.

The fifth question concerned precision. The question asked for the respondents estimation of how well the models described their processes. The rating option ranged from 1 (too specific meaning excluding paths that are possible) to 7 (too general meaning allowing paths not possible in reality).  
\begin{figure}[htb!]
	\centering
	\includegraphics[width=\textwidth]{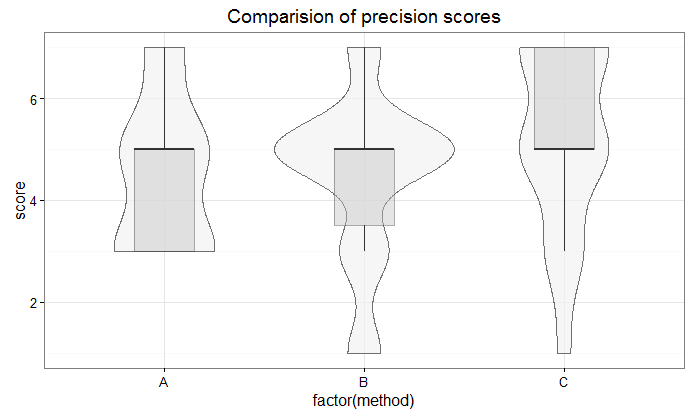}
	\caption{Results for Precision}
	\label{fig:EQ3_plot}
\end{figure}
As can be seen from figure \ref{fig:EQ3_plot}, model A and B are very similar in regards to perceived precision. The interquartile range for model A and B (both have a mean value of 4.44) are very similar whereas the results for model C show a wider distribution of responses. However, considering the distribution of the respondents results for model C (mean value of 5.22), it seems that model C is perceived to be more general as compared to model A and B.

The final question aimed at assessing perceived usefulness. As processes are often discovered for process enhancement, the respondents were asked about the usefulness of the models for improving the processes. The rating scale was from 1 (useless) to 7 (very useful). 
\begin{figure}[htb!]
	\centering
	\includegraphics[width=\textwidth]{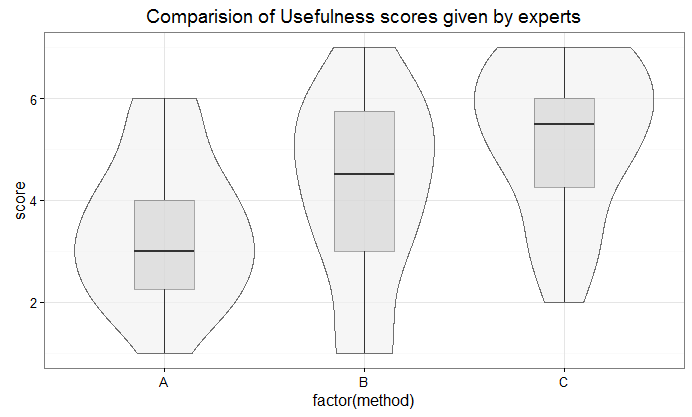}
	\caption{Results for Usefulness}
	\label{fig:EQ4_plot}
\end{figure}
The results show that model C was perceived as more useful as can be seen in Figure \ref{fig:EQ4_plot}. The difference between model A as compared with model B and C is significant. Model A has a mean value of 3.44 whereas model B has 4.17. Model C has a value of 5.22 but the difference between model B and C are less distinct. Even so, it seems that model C is perceived to be slightly more useful if it was to be used for process improvement. 

Our null hypothesis was defined as ``there is no difference in the mean rating of the models'' and the alternative was that there is at least one model that is different from the others. We reject the null hypothesis (p-value of 9.297e-07) and therefore confirm that at least one model is different. Model A is clearly different from B and C. However, we could not conclude that there are statistically significant difference between model B and C.
%

\subsection{Domain Expert Discussion Results}

Following the survey, we engaged in discussions with the domain experts about the different models. The discussions were semi-structured where a set of topical questions were asked but the discussion did not follow a strict order. The discussion questions raised were about how they perceived the models in regards to quality and understandability, how they could be improved, and if they could see any use for the models in regards to process improvement. It should be noted that usefulness of process mining techniques for discovery was not under questions. Rather, the aim of the discussions was to better understand how they perceived the generated models.

The most common observation presented by the domain experts concerned overlaying the models with additional data. Adding data about path frequency, frequency of activities, and performance data (such as different time metrics) would improve the usability and understandability of the models considerably. The reasons presented were that such data would allow distinguishing the most commonly executed paths, deviations, and clarify which paths were not taken at all (but was possible in the model). They also noted that in model C, it is possible to reach the end of the process without passing through any activity (two possible paths allow this). As such, it is needed to see which paths are actually present in the log. Such information would, when using these models for process improvement, aid in focusing efforts for further analysis.

The team leads expressed more positive sentiment in using the models for improving the processes. While they saw opportunities, the developers found the models interesting but of limited value for improving their own work. This point was quite expected as developers' work is mostly confined to one or few activities. The domain experts also expressed interest in better seeing loops in the process. The models, as they currently are, do not show where in the process loops occur. It was also noted that infrequent paths should be represented but the models could be made more simple. Understandability of the models would be improved if sub-processes would be introduced. Furthermore, the domain experts shared that the models contained more gateways and nodes than perceived necessary. In regards to the gateways, they also saw the need of annotating the gateways with numerical data such as frequency or at least default path.

The gradually gowning consensus was that the models were fairly accurate, they did capture most of the processes existing in the company, and that model c best reflected the company's everyday work. The domain experts also expressed the usefulness of such techniques for understanding the current state but wished to see simpler models (by use of sub-processes and fewer gateways) enhanced with data. It should be noted that commercial products for process discovery such as Disco \footnote{https://fluxicon.com/disco/} or Celonis \footnote{https://www.celonis.com} do provide simpler models, overlay the models with data, and allow for filtering based on activities and paths. However, open source plug-ins to ProM does not currently offer such functionality.

\subsection{Summary}

Taking all three aspects considered in the comparative evaluation, we note that model B and C are perceived as better as compared to model A. The metrics for size show that model A is clearly larger than both B and C. The number of elements (84) is clearly above the threshold suggested by research \cite{mendling2007understanding} \cite{mendling2012thresholds}. It is therefore perhaps not surprising that domain experts found this model to be the least understandable. 

\begin{table}[]
\centering
\scriptsize
\caption{Rating table}
\label{tab:rating_table}
\begin{tabular}{c|ccc|lll}
\cline{2-4}
                          & \textbf{Model A} & \textbf{Model B} & \textbf{Model C} & \\ \cline{1-4}
\textbf{Size}             & 84               & 33               & 28               & \\ \cline{1-4}
\textbf{CNC}              & 1.28               & 1.45               & 1.43               & \\ \cline{1-4}
\textbf{Density}  & 0.015               & 0.045              & 0.053              & \\ \cline{1-4}
\textbf{Understandability Q1} & 3.28                & 4.72                & 4.72                &                        &                        &                        \\ \cline{1-4}
\textbf{Understandability Q2} & 3.61                & 5.11                & 5.39                &                        &                        &                        \\ \cline{1-4}
\textbf{Understandability Q3} & 3.22                & 5.39                & 5.22                &                        &                        &                        \\ \cline{1-4}
\textbf{Correctness}  & 4.11                & 4.89                & 5.28                &                        &                        &                        \\ \cline{1-4}
\textbf{Precision}  & 4.44                & 4.44              & 5.22              &                        &                        &                        \\ \cline{1-4}
\textbf{Usefulness}  & 3.44                & 4.17              & 5.22              &                        &                        &                        \\ \cline{1-4}
\end{tabular}
\end{table}  

\begin{figure}[htb!]
	\centering
	\includegraphics[width=\textwidth]{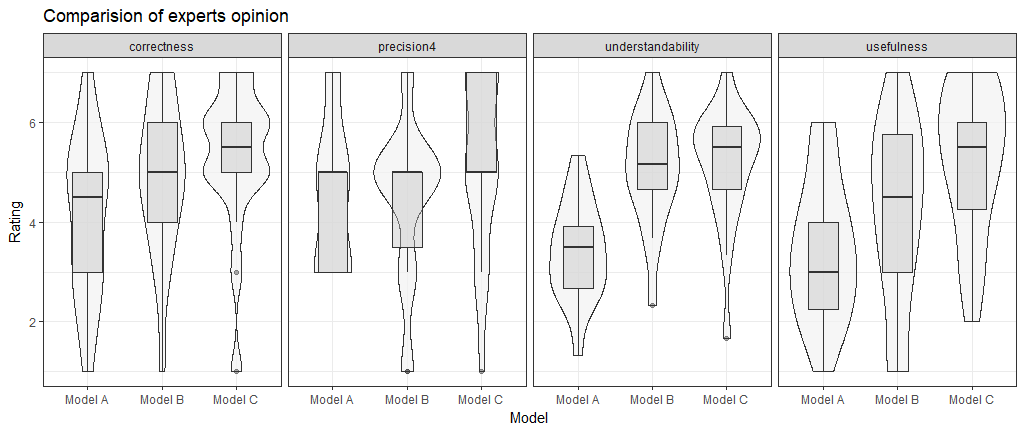}
	\caption{Comparison of Expert Opinions}
	\label{fig:Rplot.png}
\end{figure}

In regards to correctness, precision and usefulness, the domain experts clearly favor model B and C over model A. While the values for model C are slightly higher than for B, the difference is not significant enough to draw conclusions as can be seen from Figure \ref{fig:Rplot.png}. 
One might reason that the difference may be attributed to individual preference. Finally, the domain experts emphasized the benefits of having additional data captured along the models (frequency and performance metrics).

\section{Conclusion}
\label{sec:conclusion}
This paper has presented a comparative evaluation of existing implementations of automated process discovery methods using a real-life event log from an international software engineering company.
From the statistical analysis we discovered that there exist a model that is statistically different from the others, the model A. Domain experts found the Inductive Miner (model C) to be the best one, closely followed by Structured Miner (model B), but the difference is not statistically significant between these two.

Finally, domain experts found that the automated process discovery methods at current state are not as useful as they could be for their goals. Commercial packages have come further in making automatically discovered process models useful for industry. Despite the accumulated amount of research in this field, there is room for further improvements. This opens up new directions for the research, like adding frequency information to the models, splitting the models, adding tracking scales, and adding time information. In domain experts opinion these make models more usable and bring more value to them.

%
%
\bibliographystyle{splncs03}
\bibliography{references}

\end{document}